\documentclass[a4paper,11pt]{article}
\pdfoutput=1 

\usepackage{jcappub} 

\usepackage[T1]{fontenc} 

\usepackage{subcaption}
\usepackage{caption}
\usepackage[justification=centering]{caption}
\title{\boldmath RADIOACTIVE BACKGROUND FOR PROTODUNE DETECTOR}

\author[a]{M. Parvu}
\author[a,1]{, I. Lazanu\note{Corresponding author.}}

\affiliation[a]{ Faculty of Physics, University of Bucharest\\
  POBox 11, Magurele, Romania}

\emailAdd{mihaela.parvu@unibuc.ro}
\emailAdd{ionel.lazanu@g.unibuc.ro}

\abstract{The Deep Underground Neutrino Experiment (DUNE) is a leading-edge, international experiment for neutrino science and proton decay studies.  This experiment is looking for answers regarding several fundamental questions about the nature of matter and the evolution of the universe: origin of matter, unification of forces, physics of black holes. Two far detector prototypes using two distinct technologies have been developed at CERN. The prototypes are testing and validating the liquid argon time projection chamber technology (LArTPC). In neutrino physics, as well as in any experiment with rare interaction rate, the good knowledge of the radioactive backgrounds is important to the success of the study.

Muons and neutrons represent the main sources of background for this kind of experiments. In this paper, we have considered two sources of neutrons: cosmic neutrons and neutrons coming from the accelerating tunnel. Also, cosmic muons are taken into account. The contribution of these particles to the production of radioactive isotopes inside the active volume of the detector in comparison to the one corresponding to muons is shown. Also, simulations of nuclear reactions for the processes of interest for investigating the radioactive background due to the lack of measurements or insufficient experimental data are presented. Most of the results presented in this paper will be of interest for the future underground DUNE experiment.}

\begin{document}
\maketitle
\flushbottom

\section{Introduction}
In the next generation of neutrino experiments, DUNE (Deep Underground Neutrino Experiment) will represent an ensemble of unique parameters: the most intense neutrino beam, a deep underground site and massive detectors using LAr as active medium and with two distinct TPC technologies: Single-phase (SP) and Double Phase (DP) \cite{Abi:2018dnh}. The Single Phase technology, now at mature stage, started with the pioneered ICARUS project \cite{Rubbia:1977zz}. For the Dual Phase the first large scale application was the WA105 DP demonstrator and presently, after more R$\&$D stages, ProtoDUNE-DP detector is in tests. Currently the Double Phase concept was replaced by the vertical drift option, where two separate drift volumes are defined by a cathode plane at mid height in liquid argon. This new design represents an important evolution of the detector, but does not change the physics aspects discussed in the present paper.

The primary science goals of DUNE include: a comprehensive program of neutrino oscillation studies, search for proton decay, neutrino physics of supernovae, as well as other accelerator-based neutrino flavor transition measurements with sensitivity to beyond the standard model (BSM) physics.
Also, DUNE could contribute to measurements of solar neutrinos. DUNE has potential to record an enormous solar neutrino contribution \cite{Capozzi:2018dat}, with sensitivity to $^8$B decay as well as \textit{hep} solar processes, considering CC and elastic scattering, but depending on the radioactive background. As these processes are in the same energy range as the radioactive background, we are motivated to explore the details of these backgrounds with simulations.

An increased background rate will affect physics sensitivity to low-energy neutrinos and thus identification of all sources of unwanted processes are essential. Also if $^{222}$Rn and its progenitors or neutrons have a high rate, this would make the $\phi(hep)$ measurement inconclusive. At present, for ProtoDUNE and DUNE the radioactive background is still uncertain due to the lack of complete measurements considering complex fields of radiation and full simulations. For the ground conditions as in the case of ProtoDUNE detectors, neutrons and secondary particles resulting from their interaction represent the main source of background. Also, uncertainties in neutron transport exist and must be considered. For steel of the structure and stainless steel in the membrane of the cryostat exist some uncertainty in the concentration of U/Th, but also  uncertainty in the  cross-sections of neutrons in the elements of the composition exist. More importantly, measurements of the spatial and vertical gradient of radon levels could be made if these abundances could be reduced. In the DUNE case, the four detector modules will be installed approximately 1.5 km underground and the processes that will contribute to radioactive background will be different. A part of the cosmic rays will reach underground but with a different energetic flux but the cross sections remain of interest.
\newline
The paper is organized as follows. In Section \ref{sec:EHN1} some aspects of the ProtoDUNE-DP detector in EHN1 hall are discussed. The radioactive background due to cosmic and beam neutrons as well as due to cosmic muons are analized in Section \ref{Backgr}. The results of the simulations and phenomenological estimations are the subject of the next section. In Section \ref{argon} the production mechanisms of unstable argon isotopes or other radioactive nuclei in bulk LAr detector with neutrons are discusssed. In the last section the conclusions are presented.

\section{ProtoDUNE in EHN1}
\label{sec:EHN1}

In CERN’s North Area Experimental Hall, the EHN1 hall is 70 m in extent. The detector prototypes (NP-02 and NP-04) are installed in the extension part of the EHN1 building – for details see for example Figure 1 of the paper of Charitonidis and Efthymiopoulos \cite{Charitonidis:2017omo}. Besides running with cosmics, the two detector prototypes will also receive beams which will induce a supplementary background. The main beam design parameters necessary for the experimental proposals consists in the following particles: $\pi^{+,-}$, $\mu^{+,-}$, $e^{+,-}$, K, p with parameters: $\Delta$p/p $<$ 5$\%$, beam size rms $\sim$10 cm at the entrance of the cryostats and the maximum rate 100  Hz. The momentum is $<$ 7 GeV/c for protoDUNE-SP and < 12 GeV/c for ProtoDUNE-DP respectively \cite{Charitonidis:2017omo}.

\begin{figure}[h]
    \centering
    \includegraphics[scale=0.6]{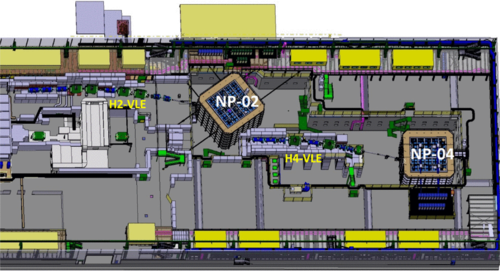}
    \caption{EHN1 hall layout taken from \cite{Charitonidis:2017ffh}.}
    \label{ehn1}
\end{figure}

A special case is ProtoDUNE-DP detector because of the vicinity of the beam used for ProtoDUNE-SP. This peculiarity of the geometrical configuration is put in evidence in the 3D model layout of the H4-VLE beam line presented in Figure 6 of the cited paper. Under these circumstances H4-VLE beam will produce a supplementary background in the cryostat of the double phase detector. These contributions are also considered in this article. A schematic layout of the geometry is presented in Figure \ref{ehn1}.
 
Even though the accelerating pipe is shielded using concrete blocks, there still exists an emerging flux of neutrons with a wide energy spectrum \cite{Bilski:2006nv} shown in Fig. \ref{beam-neutroni} which is considered as input for simulation of the reactions in the bulk of the detector and its construction elements. Given the building layout where the detectors are placed, the localization of the ProtoDUNE-DP with respect to the beam pipe can be seen.

\section{Background due to neutrons and muons}
\label{Backgr}

Neutrons are an important background in low rate physics experiments. Because of the lack of electrical charge, these particles can cross long paths suffering just elastic and inelastic collisions. Even if neutrino experiments are usually placed underground, there can still exist neutrons produced by muon-induced reactions, in the decay chains of U and Th and possible residual moderated neutrons from cosmic rays. In accelerated beams, neutrons can appear as secondary particles due to beam interaction in the accelerating tunnel and this component must be taken into account.
These particles are unwanted because they can simulate a signal event corresponding to low-energy neutrinos, as well as for proton decay and dark matter observation. Moreover, they can produce long-lived radioisotopes and secondary gamma rays processes in the detector or its shielding.

\subsection{Argon, its isotopes and intrinsic sources of contamination}

Argon is an abundant element, present in Earth's atmosphere. A great number of argon isotopes exist, between which three are stables: $^{40}$Ar (99.6003\%), $^{36}$Ar (0.3365\%),$^{38}$Ar (0.0632\%) (in the parenthesis their abundances are shown);  other isotopes are unstable, but only three are long lived radioactive ($^{37}$Ar, $^{39}$Ar and $^{42}$Ar) and one relatively long half life, around 100 minutes ($^{41}$Ar). All other are irrelevant because of their very short half-life and thus no effects are expected. 
Except $^{37}$Ar isotope that decay via EC in stable $^{37}$Cl, others decay via $\beta$ emission in potassium isotopes and, as has been discussed in \cite{Parvu:2017xde} can perturb the ionization and scintillation signals. Supplementary, contamination with natural $^{40}$K can exist.

\subsection{Cosmic neutrons and muons}

In three successive papers, Ziegler \cite{Ziegler1, Ziegler2, Ziegler3} investigated the cosmic ray intensities at sea level starting from experimental measurements taken at different locations and at different altitudes, all normalized for New York City. For the present discussion, of interest are only the results for neutrons and muons respectively. Using data obtained in the range of 10 MeV up to 10 GeV, Ziegler obtained a fit and an analytical formula for neutron flux intensity that reproduce data within 1\% measurements precision. In these calculations no correction for solar cycle is considered, because the scatter of data is much larger than the effects of the solar cycle. For other locations the cosmic ray flux is evaluated on the basis of the latitude, longitude, and altitude; using the experimental values for the change in the neutron flux with altitude, and cascade-calculation mean attenuation lengths for the other particles. For CERN (Geneva) the correction factor is 1.43 relative to New York.

\begin{figure}[h]
    \centering
    \includegraphics[scale=0.45]{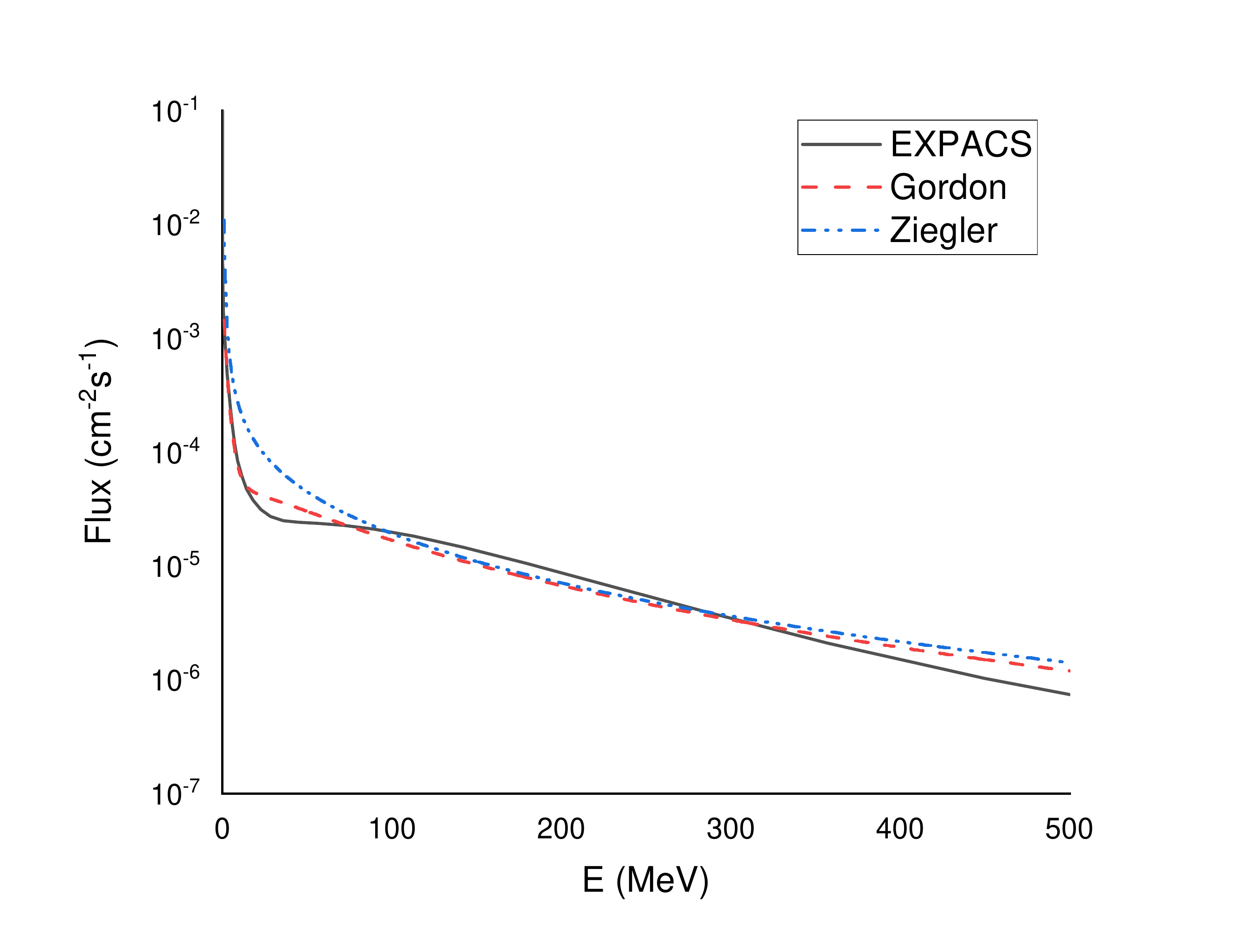}
    \caption{Neutron flux using EXPACS code \cite{10.1371/journal.pone.0144679} and the parameterizations of Ziegler \cite{Ziegler3} and Gordon \cite{Gordon2005}.}
    \label{neutroni}
\end{figure}

Gordon and co-workers \cite{Gordon2005} obtained other measurements of the neutron flux and its energy distribution as produced by cosmic rays at several locations in the US using Bonner sphere spectrometers. The data cover over twelve decades of neutron energy, from meV to GeV. In addition, an analytic expression fits the neutron spectrum above about 0.4 MeV. Their results are limited to neutrons. In Figure \ref{neutroni} the energy dependence of the neutron fluxes in accord with the parametrizations of Ziegler and Gordon are shown.

In order to minimize the need for time-consuming MC simulation of the cosmic-ray propagation, Sato \cite{10.1371/journal.pone.0144679}, based on a comprehensive analysis of the simulation results, proposed an analytical model for estimating the atmospheric cosmic-ray spectra for neutrons, protons, He nuclei, muons, electrons, positrons and photons applicable to any global conditions. This model is called "PARMA": PHITS based Analytical Radiation Model in the Atmosphere and is implemented in EXPACS: "EXcel-based Program for calculating Atmospheric Cosmic-ray Spectrum".
In their model Sato determine the numerical values of the parameters of the analytical functions using the results of the extensive air shower simulations. Neutron fluxes obtained using the parametrizations of Ziegler and Gordon. as well as the results of EXPACS code, are included in Figure \ref{neutroni}. A very good agreement can be observed excepting the 10 -70 MeV energy interval. 

\begin{figure}[h]
    \centering
    \includegraphics[scale=0.45]{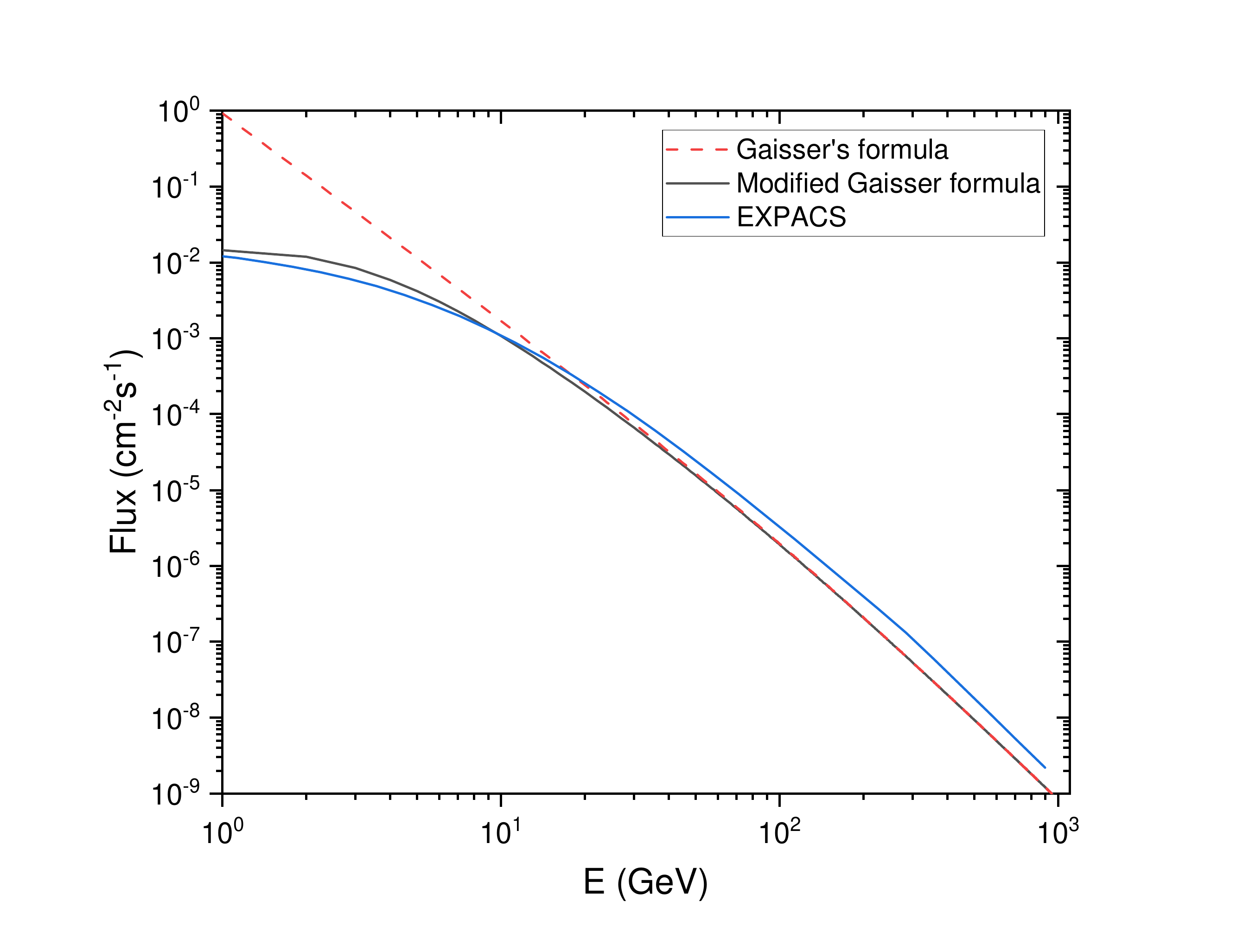}
    \caption{Muon flux at sea level using Gaisser's parametrization \cite{Gaisser}, modified Gaisser formula \cite{Guan:2015vja} and EXPACS results \cite{10.1371/journal.pone.0144679}.}
    \label{muonflux}
\end{figure}

Parametrizations of the energy dependence of the muon flux using two parametrizations and EXPACS are indicated in Figure \ref{muonflux}. Since at low energies, muon decay should be considered, the standard Gaisser’s formula should be modified in order to describe better the experimental results. 


\subsection{Beam induced neutrons}

\begin{figure}[h]
    \centering
    \includegraphics[scale=0.45]{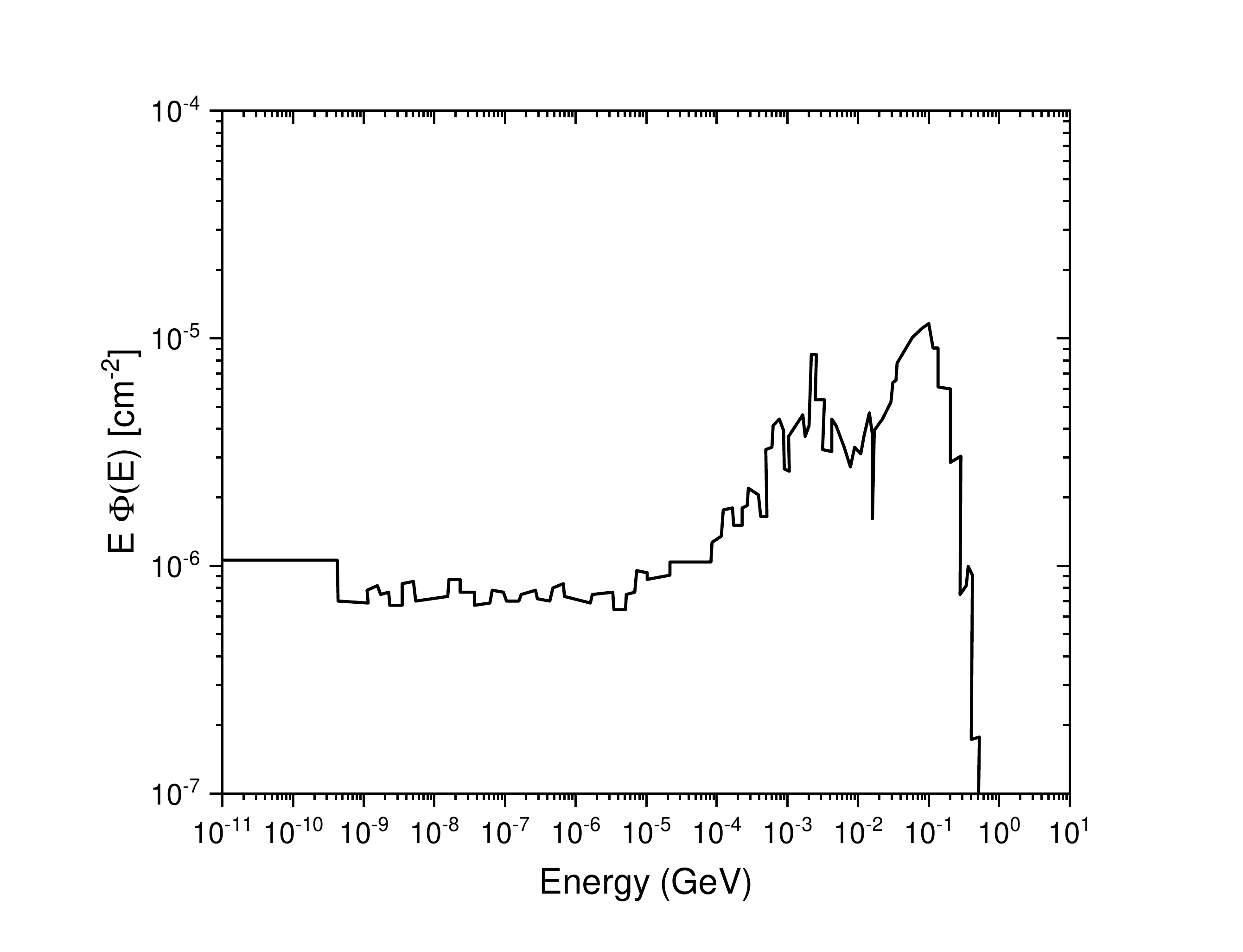}
    \caption{Beam induced neutron flux adapted from \cite{Bilski:2006nv}.}
    \label{beam-neutroni}
\end{figure}

The neutron energy distribution per primary incident beam particle calculated with the FLUKA Monte Carlo code by Bilski et.al. \cite{Bilski:2006nv} is shown in Fig. \ref{beam-neutroni} and it is used, after normalization, for calculation of the doses. In this case is considered that a positive hadron beam (35\% protons, 61\% pions and 4\% kaons) with momentum of 120 GeV/c is stopped in a copper target, 7 cm in diameter and 50 cm in length.

The new beam line for ProtoDUNEs is designed to produce a positive low-energy hadron beam, consisting of 70\% $\pi^+$, 25\% p and 5\% K$^+$ with $\sim$1\% $\Delta$p/p momentum spread, and with an intensity of 10$^6$ particles per pulse \cite{Charitonidis:2016enz}. This intensity is supposed to be identical with the case considered in this paper. The neutron energy spectrum is normalized to the proton beam that interacts in the copper target.

The spectrum outside the iron shield is dominated by neutrons in the 0.1–1 MeV range while the energy distribution outside both concrete shields shows a distribution in which this energy region is reduced while the contribution of 10–100 MeV neutrons dominates. It is supposed that the experimental configuration along the beam for ProtoDUNE-SP detector is similar to the one discussed in reference \cite{Bilski:2006nv}. The fluence rate of other hadrons is much lower than that of neutrons. 

\subsection{Muon induced neutrons}

The production of fast neutrons from cosmic-ray muons was measured or predicted by different authors and groups, see for example Boehm et al. \cite{Boehm:2000ru} or  Wang et al. \cite{Wang:2001fq}. Fast neutrons from cosmic-ray muons are produced by the following processes: muon capture,  muon spallation processes (interactions with nuclei via a virtual photon and nuclear disintegration), elastic scattering with neutrons from nuclei, photonuclear reactions. 
The different types of results (measurements, theoretical models, empirical formulae and simulations) are contradictory in the literature. A very interesting analysis of the dependencies and possible parameterizations of the neutron spectra was done by Wang and co-workers. 

 \textbf{Neutron yield.}  All neutrons, either primary or secondary, are included in the calculations. Wang and collaborators suggest that the neutron yield per muon, scalable at different depths (water equivalent), can be fit as:

\begin{equation}
N_n=4.14 \times 10^{-6} E[\rm GeV]_{\mu}^{0.74} \textrm{neutrons/(\rm $\mu$ g cm$^{-2}$)}.
\end{equation}

 Authors consider the origin of the neutrons direct muon spallation, photonuclear disintegration, neutron spallation, proton spallation, pion spallation and capture (for negative pions) and others.

 For muon energies between 3.9 and 270 GeV, the neutron energy spectrum can be fitted with the universal empirical function: 

\begin{equation}
\frac{dN}{dE_{n}}=A\Big[\frac{e^{-7E_n}}{E_n}+(0.52-0.58e^{-0.0099E_{\mu}})e^{-2E_n}\Big].
\end{equation}

Here, A is a normalization factor.

The second quantity is the neutron multiplicity (M), for which exists a universal empirical parameterisation:

\begin{equation}
\frac{dN}{dM}=A \Big[{e^{-A(E_{\mu})M}+B(E_{\mu})e^{-C(E_{\mu})M}} \Big],
\end{equation}

 The quantities A, B and C are obtained as: 

\begin{equation}
\begin{aligned}
&A(E_{\mu})=0.085+0.54e^{-0.075E_{\mu}};\\
& B(E_{\mu})=\frac{27.2}{1+7.2e^{-0.076E_{\mu}}};\\
& C(E_{\mu})=0.67+1.4e^{-0.12E_{\mu}}.
\end{aligned}
\end{equation}

 For the muon energy range between 40 GeV up to $\sim$ 400 GeV, Malgin and co-workers obtained an universal but empiric parameterization relative similar for muon induced neutron yield and thus neutron induced rate \cite{Agafonova:2013bva,Malgin:2017dwh}. In  the general case, for a muon with an energy $E_{\mu}$, the neutron yield in a material with mass number A is:

\begin{equation}
Y_{n}(A,E_{\mu})=\frac{N_A}{A} \langle \sigma_{\mu A} M \rangle =\gamma E_{\mu}^{\epsilon 1}[GeV]A^{\epsilon 2},
\end{equation}

where N$_A$ is Avogadro’s number, $\langle \sigma_{\mu A}$M$\rangle$ is a mean value of the product of a $\mu$ + A interaction cross-section and neutron multiplicity, M. In this equation, neutron yield is measured in (n/$\mu$/(g/cm$^2$)). Here the constants have the following values: $\gamma$=$4.4 \times 10^{-7}$ cm$^2$g, ${\epsilon 1}$= 0.78, ${\epsilon 2}$= 0.95.

 In accord with Agafonova and Malgin the neutron yield in iron is in the range $1.6 \times 10^{-4}$ to $2.06 \times 10^{-3}$, $2.4 \times 10^{-5}$ to $2.99 \times 10^{-4}$ for polyurethane foam and $1.2 \times 10^{-4}$ to $1.5 \times 10^{-3}$ for Ar. These results argue that muons do not contribute significantly to increasing of the neutron flux for the energy range considered.

 Rapid neutrons from cosmic-ray muons are produced in interactions with nuclei as elastic scattering, muon spallation, photo-nuclear reactions or as secondary particles after these primary processes.

 Neutrons can be also produced after nuclear muon capture. This process is reasonably well understood but their contribution to background is not usually considered. These neutrons can be produced after muon capture in atmospheric nuclei, in components of the cryostat (in polyurethane foam,  in the stainless steel membrane, as well as in bulk of LAr).

 \textbf{Negative muon capture in nuclei}. The negative muon is captured in an atomic orbit when it is stopped in matter. It then cascades to the 1S level where it either decays or is captured by the nucleus following the nuclear reaction $\mu^-+(A,Z) \rightarrow (A^*,Z-1)+\nu_{\mu}$. 
Around 100 MeV (most of the energy released) is carried away by the neutrino. The mean excitation energy of A$^*$ is around 15 to 20 MeV; thus A$^*$ can de-excite by emitting one or more neutrons, or charged particles, or it may de-excite via the ordinary electromagnetic mode. Studies and measurements  of Wyttenbach  and collaborators for nuclei between Na up to Bi \cite{Wyttenbach:1978rp} suggest the following conclusions:
The dominant reaction is $(A,Z)(\mu^-,xn)(A-x,Z-1)$ with one or more neutrons in the final state (x=1 or higher). These channels and electromagnetic de-excitation mode account for more than 95\% of the total reaction probability. Combined with channels with charged particles in the final states, Wyttenbach and and co-workers reported the relative probabilities for $(\mu^-,pxn)$ reactions for a given target vary as 1:6:4:4 for x=0, 1, 2, 3.

 From the theoretical point of view, the probability W of each type of reaction depends on the atomic number of the target and can be described by W = a exp(-bV), where V is the Coulomb barrier of the compound nucleus for the ejected charged particle with the factor b the same for all $(\mu^-,pxn)$ reactions and similar for $(\mu^-,\alpha)$ reactions.

In Table \ref{reaction_probability} we estimated the corresponding probabilities for some targets of interest as interpolated experimental data using the curves published by Wyttenbach and collaborators.

\begin{table}
\begin{center}
\begin{tabular}{ |c|c|c|c|c|c| }  
\hline
 \hline
Nucleus & $(\mu^-,p)$ & $(\mu^-,pn)$ & $(\mu^-,p2n)$ & $(\mu^-,p3n)$ & $(\mu^-,\alpha)$\\
 \hline
 Argon   & $4.5 \times 10^{-3}$ & $2.5 \times 10^{-2}$ & $2 \times 10^{-2}$ & - & $4 \times 10^{-2}$\\
 Iron & $2.5 \times 10^{-3}$ & $1.4 \times 10^{-2}$ & $1.2 \times 10^{-2}$ & $1.1 \times 10^{-4}$ & $1 \times 10^{-3}$\\
 Chromium & $2.5 \times 10^{-3}$ & $1.5 \times 10^{-2}$ & $1.4 \times 10^{-2}$ & $1.3 \times 10^{-4}$ & $1.6 \times 10^{-3}$\\
  Nickel & $2.4 \times 10^{-3}$ & $1.4 \times 10^{-2}$ & $1 \times 10^{-2}$ & $1 \times 10^{-4}$ & $8 \times 10^{-4}$\\
 \hline
\end{tabular}
\caption{Muon-induced reaction probabilities for different processes in argon and in the major constituents materials of the cryostat.}
\label{reaction_probability}
\end{center}

\end{table}

 A supplementary background is expected due to muon spallation. A detailed study was published in Ref. \cite{Zhu:2018rwc} for argon, considering the production of unstable isotopes, sources for later beta background. 

\section{Simulation results}

 We have considered the investigation of the radioactive isotope production inside the active volume and the deposited energy in the detector.
The simulations were done using FLUKA Monte Carlo code \cite{Bohlen:2014buj,Ferrari:2005zk} which can be used for calculations of particle transport and interactions with matter. The code covers a wide particle range more than 60 different particles: photons and electrons (with energies from 1 keV to thousands of TeV), neutrinos, muons, hadrons (energies up to 10 PeV), the corresponding antiparticles, neutrons and heavy ions.  Microscopic models are implemented with priority, consistency among the reaction steps and reaction types is ensured, as well as the respecting of the conservation laws at each step. The code allows the estimation of different quantities like energy density, energy and momentum transfer in a geometry-independent binning structure (cartesian or cylindrical), energy deposition, fluence and current as a function of energy and angle, particle yields or differential cross sections, residual nuclei etc. Another important feature is that very complex geometries can be implemented, due to the Combinatorial Geometry package.

 In order to simulate the muon and neutron interactions, a simplified but realistic geometry of the detector was implemented. 
The main target was represented by a volume of liquid argon (854.8 $\times$ 790 $\times$ 854.8 cm$^3$), surrounded by 80 cm of polyurethane foam (C$_{27}$H$_{36}$N$_{2}$O$_{10}$) and 3 mm of stainless steel membrane (73\% Fe, 18\% Cr, 9\% Ni). Naturally occurring iron consists of four stable isotopes: 5.845\% of $^{54}$Fe, 91.754\% of $^{56}$Fe, 2.119\% of $^{57}$Fe and 0.286\% of $^{58}$Fe. The detector was finally placed in dry air at 20 $^{\circ}$C. In the case of the materials the air was predefined in the code (dry air at 20 $^{\circ}$C), whereas the liquid argon, polyurethane and stainless steel were user-defined materials. Due to the energies in the neutron spectra the activation of the LOW-MAT card for each element was needed, in order to activate the transport of low energy neutrons.

\subsection{Deposited doses and other characteristics due to neutrons in bulk cryostat of detector}

\begin{figure}[h]
    \centering
    \includegraphics[scale=0.6]{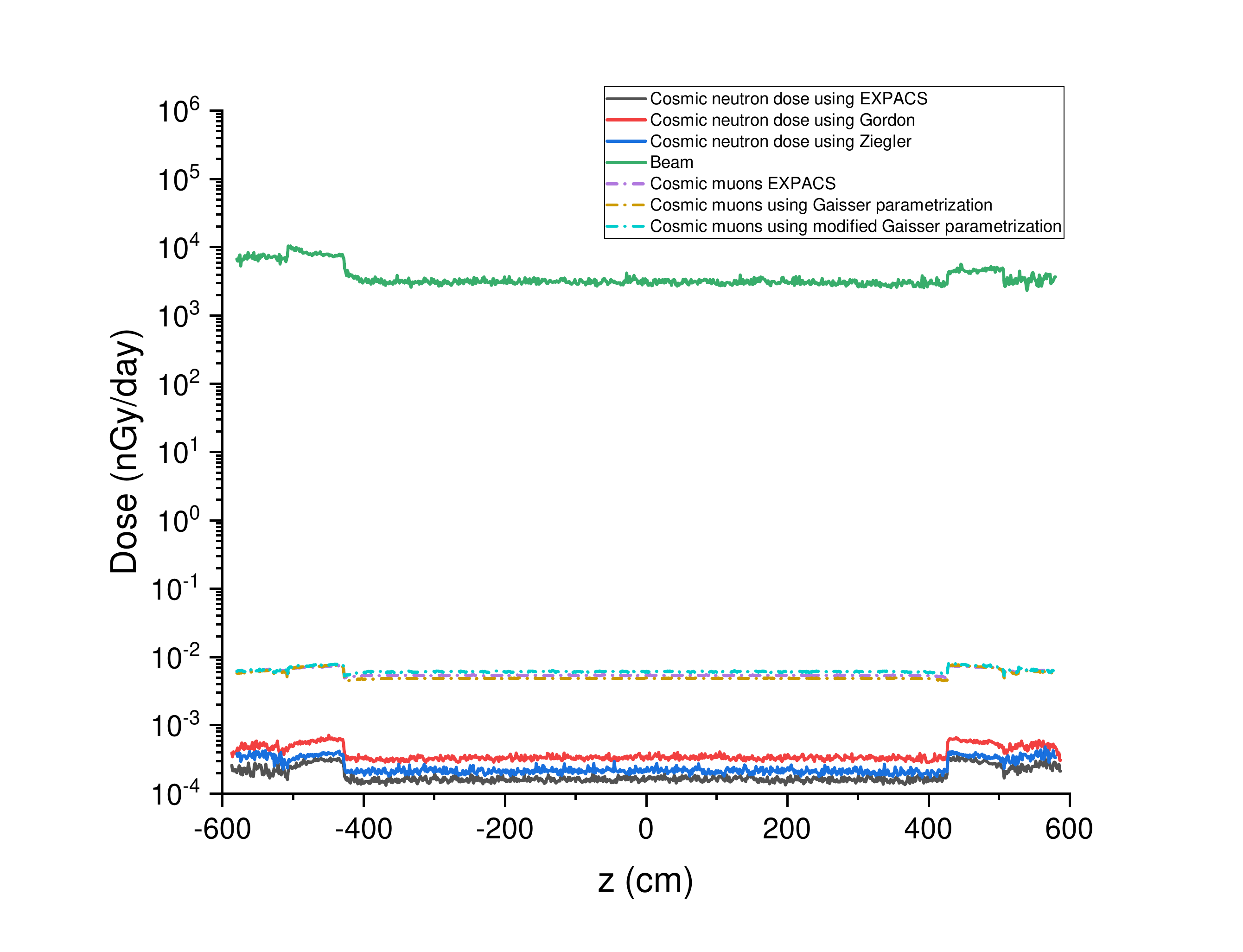}
    \caption{Daily doses induced by cosmic neutrons and beam induced neutrons in ProtoDUNE-DP.}
    \label{dose}
\end{figure}

 In Figure \ref{dose} the daily doses deposited by muons, cosmic neutrons and beam induced neutrons in ProtoDUNE-DP as a function of the position in the cryostat are presented. In the simulated values the presence of the metallic external structure, the polyurethane foam layer, and the membrane are clearly put in evidence. 
For the beam-induced neutrons the asymmetry in the deposited energy due to the existing geometry is observable. 

The dose contribution from neutrons is around (2 - 3) $\times$ 10$^{-4}$ nGy/day considering any parametrization of the flux. Muons contribute with a dose around 5 $\times$ 10$^{-3}$ nGy/day. The main background source is due to beam induced neutrons.

\begin{figure}
 \begin{subfigure}{.5\textwidth}
 \centering 
  \includegraphics[scale=0.6]{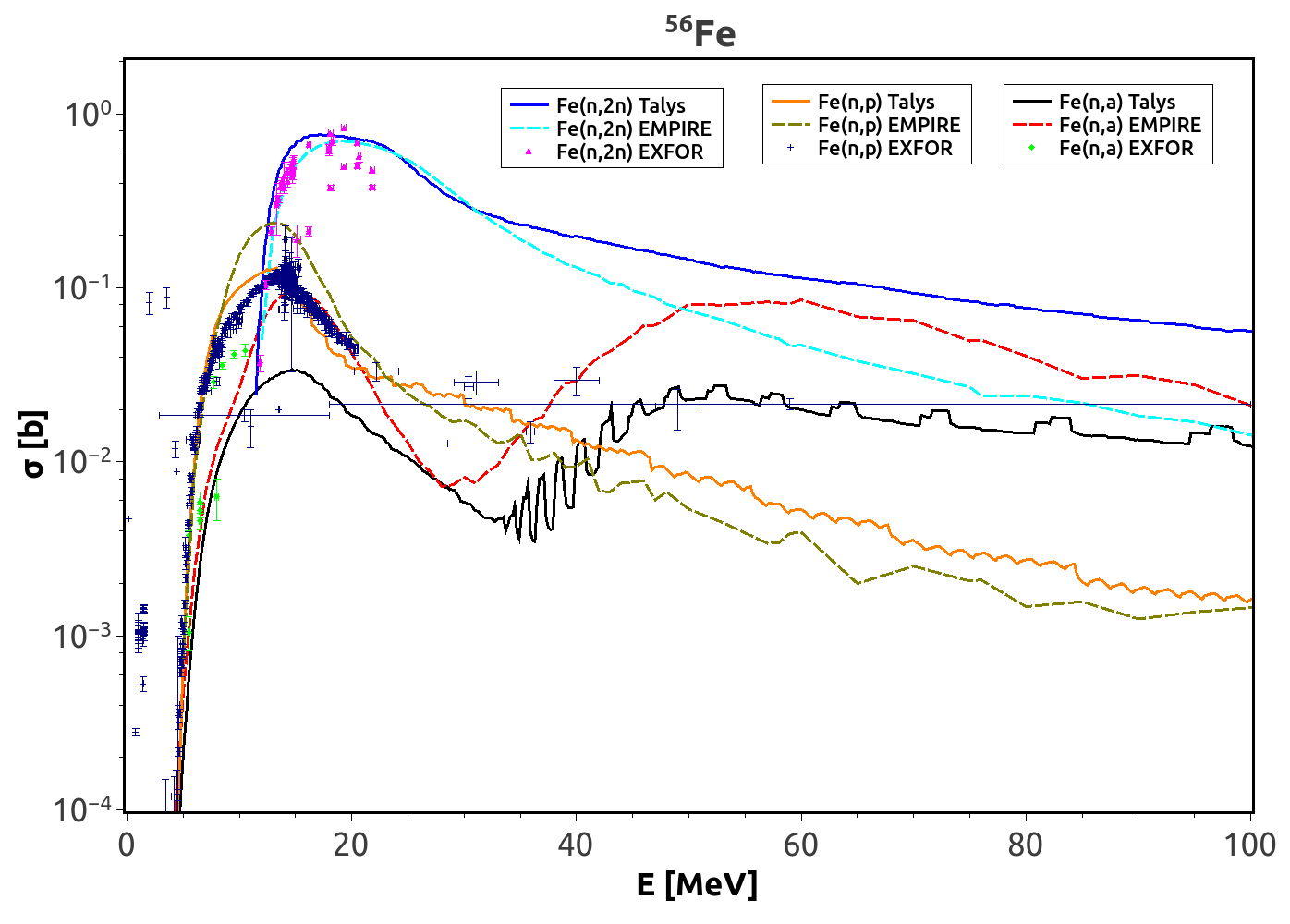}
    \caption{}
    \label{Fe}
 \end{subfigure}
 \begin{subfigure}{.5\textwidth}
\centering  
  \includegraphics[scale=0.6]{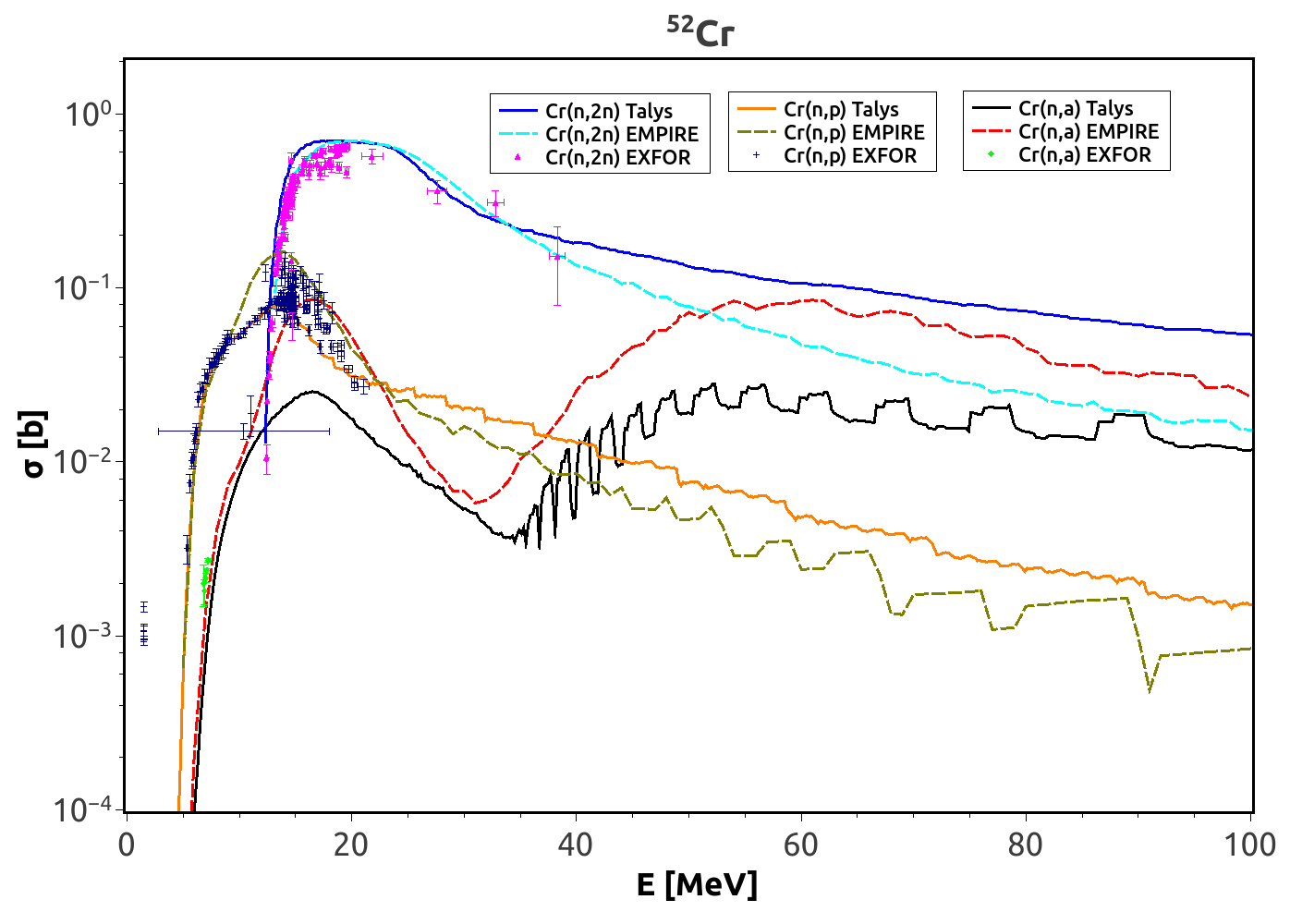}
    \caption{}
    \label{Cr}
   \end{subfigure}

   \centering   
   \begin{subfigure}{.5\textwidth}
\centering    
    \includegraphics[scale=0.6]{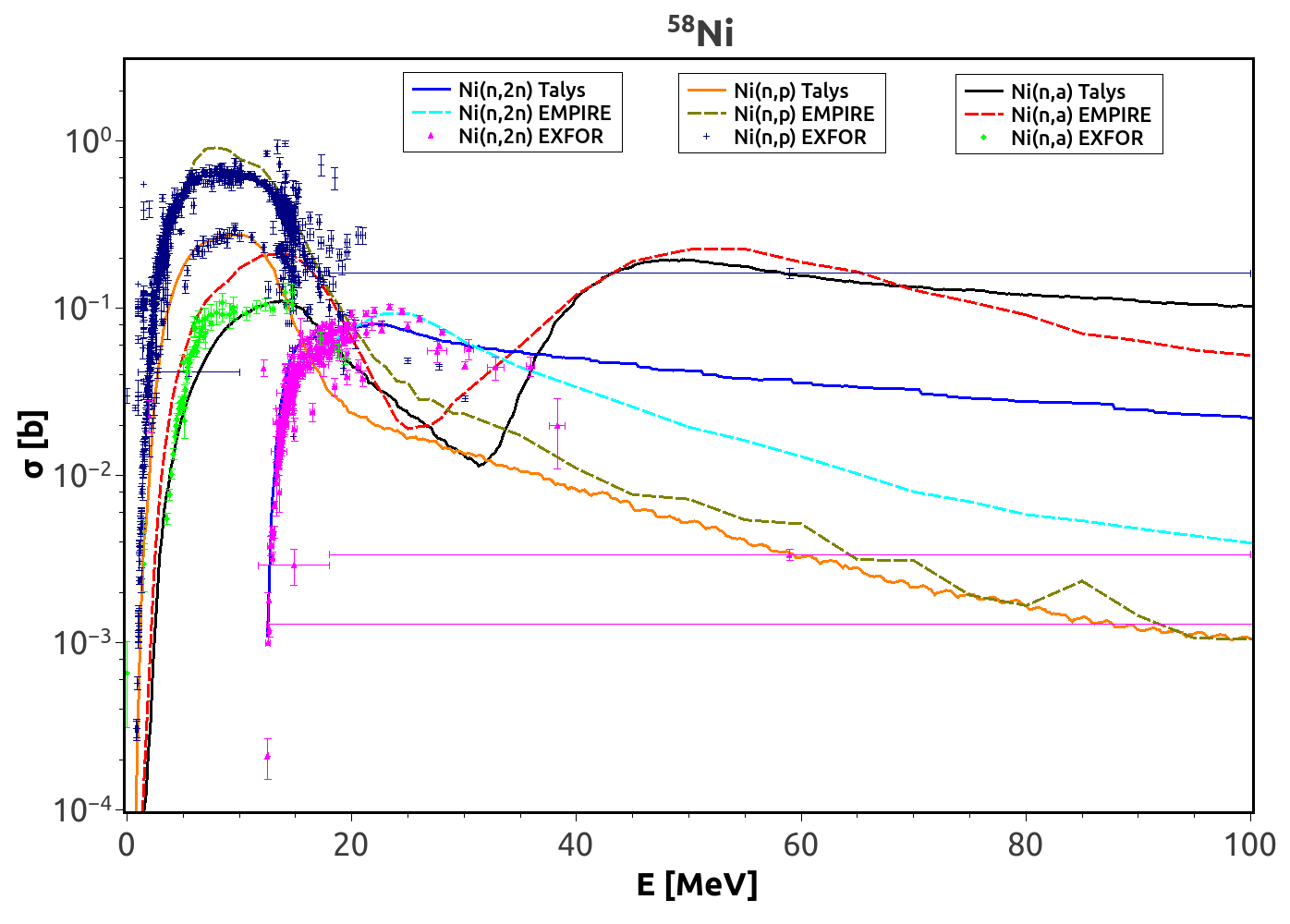}
    \caption{}
    \label{Ni}
    \end{subfigure}
    \caption{The energy dependence of the cross sections for the reactions  (n,2n), (n,p) and (n,$\alpha$) in $^{56}$Fe, $^{52}$Cr and $^{58}$Ni. Experimental data are reproduced from EXFOR \cite{Otuka:2014wzu} and predicted cross sections are obtained using Talys \cite{Koning:2019qbo} and EMPIRE \cite{Herman:2007}.}
    \label{Sectiuni}
\end{figure}

 In Figure \ref{Sectiuni} a), b) and c) are presented the cross sections of the processes (n,2n), (n,p) and (n,$\alpha$) in $^{56}$Fe, $^{52}$Cr and $^{58}$Ni, the dominant elements in the composition of the stainless steel as membrane of the cryostat of detector. In these cases all processes are possible sources of background on the inner surface of the membrane. Experimental data for energy dependencies of the cross sections are obtained from EXFOR \cite{Otuka:2014wzu} database and  calculated cross sections are obtained using Talys \cite{Koning:2019qbo} and EMPIRE \cite{Herman:2007} codes.

 In the composition of the membrane, iron is the major element in the composition next to chromium and nickel.
For the reaction (n,2n) in iron ($^{56}$Fe) experimental data exist in the energy range 10-20 MeV and these are relatively well reproduced by Talys and EMPIRE codes. At energies above 30 MeV discrepancies between the two codes appear and these differences increase with energy. In the case of (n,p) reaction, both of the codes predict a maximum in the cross section at the same incident energy. For the process with alpha particle in the final state the models are not able to reproduce the experimental data available. The comparison between data and predicted values using considered codes is presented in Figure \ref{Sectiuni} a).

 In the case of chromium, for the reaction (n,2n) the data are very well reproduced by Talys and well by EMPIRE. Data are available up to 37 MeV.  In the case of (n,p) reaction experimental data have a large dispersion for the same energy and thus these values are similar between Talys and EMPIRE predictions. All these reactions are presented in the Figure \ref{Sectiuni} b).

 Nickel is the second major element constituent of the  stainless steel. For all three reactions considered the codes predict a maximum in the cross sections (values between 0.1 -1 barn) for energies in the range of 5-15 MeV. Reaction (n, 2n) is very well reproduced both Talys and EMPIRE. For the reaction (n,p) two sets of experimental data with major discrepancies in the energy region 5-10 MeV exists, every code reproduce only a data set. The experimental data for the reaction (n, $\alpha$) are close between the predicted values of both codes. All these results are  reproduced in Figure \ref{Sectiuni} c).

 Thus, if these reactions are produced inside the membrane, at the contact with the liquid argon, there will exist free neutrons, protons and alpha particles that will produce secondary reactions. For charged particles this region is defined by their range for corresponding energies. In the next figures (Fig. \ref{prange}, \ref{rangealpha}) energy dependencies of the  ranges of protons and alpha particles in LAr  are represented. For protons the reference \cite{Acciarri:2013met} is used and for alpha particles calculations are done in the frame of SRIM \cite{SRIM} and FLUKA codes. In the neutron case the thermalization process was not considered. 

\begin{figure}
 \centering 
  \includegraphics[scale=0.55]{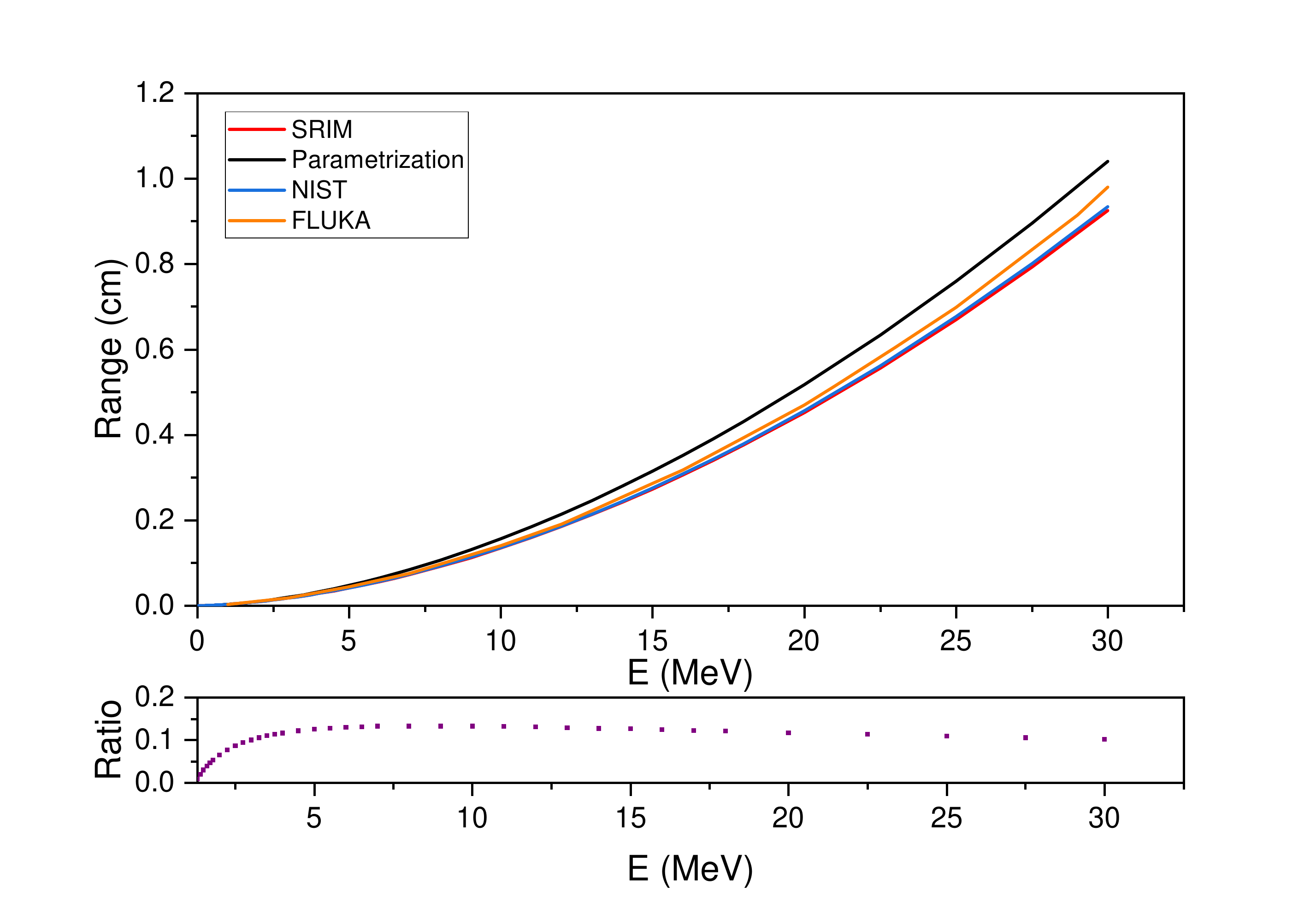}
    \caption{Proton range in LAr using SRIM \cite{SRIM}, FLUKA \cite{Bohlen:2014buj,Ferrari:2005zk}, NIST \cite{NIST} and the parametrization of ArgoNeuT Collaboration \cite{Acciarri:2013met}. Relative value between maximum (Parametrization) and minimum (SRIM) range values vs. energy is presented.}
    \label{prange}
\end{figure}

\begin{figure}
\centering  
  \includegraphics[scale=0.35]{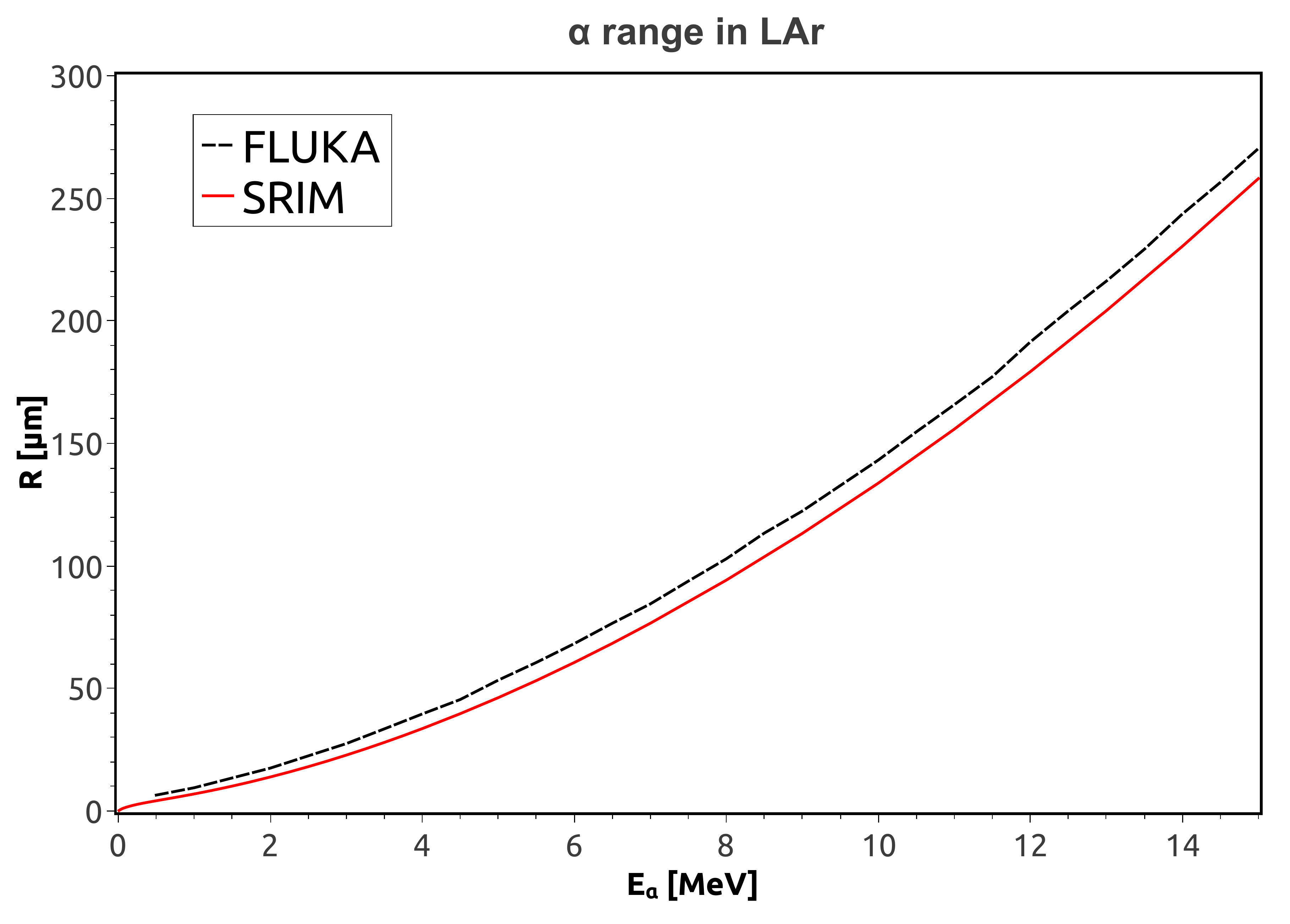}
    \caption{Range of alpha particles in LAr. The calculated values are obtained using FLUKA and SRIM codes.}
    \label{rangealpha}
\end{figure}

\section{ Production mechanisms of unstable argon isotopes or other radioactive nuclei in bulk LAr detector due to neutrons}
\label{argon}

 The primary product of the  $\mu$ capture in Ar is a Cl* nucleus excited from zero to about 100 MeV. This state de-excites then in two ways: electromagnetic transitions (if the excitation energy is less than about 6 MeV) or emission of one or several nucleons.

 The reactions (n,2n) and (n,2n$\gamma$) represent the direct and dominant nuclear reactions to produce $^{39}$Ar if the process is on $^{40}$Ar isotope or on 
$^{38}$Ar(n,$\gamma$)$^{39}$Ar, or in sequential processes (n,np), (n,d) and (n,d$\gamma$) on $^{40}$Ar followed by the beta decay of $^{39}$Cl. This way is disfavored because half-life of $^{39}$Cl is only on 55.6 minutes. 

\begin{figure}[h]
    \centering
    \includegraphics[scale=0.8]{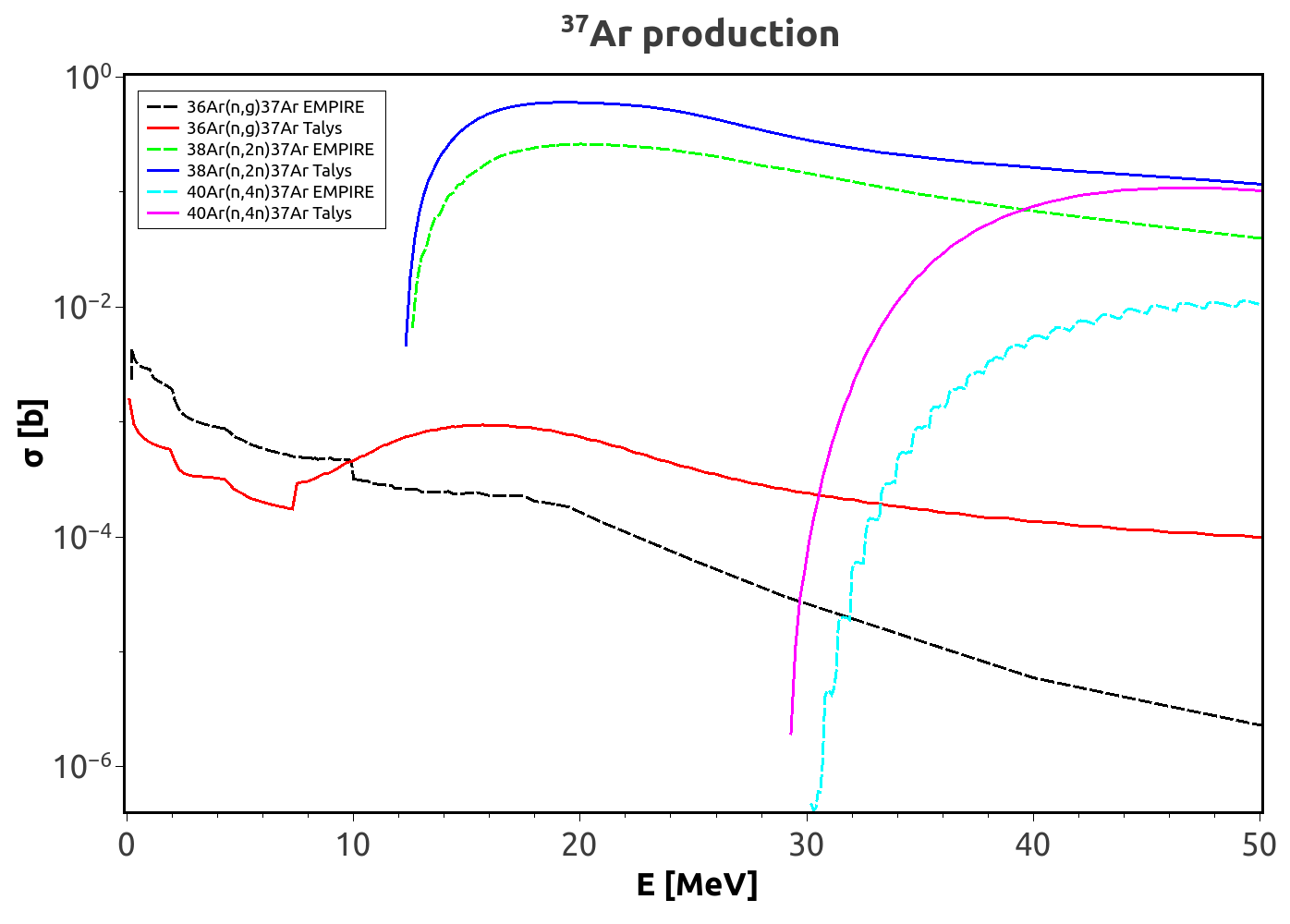}
    \caption{The energy dependence of the cross sections for the $^{36}$Ar(n,$\gamma$)$^{37}$Ar, $^{38}$Ar(n,2n)$^{37}$Ar and $^{40}$Ar(n,4n)$^{37}$Ar reactions.}
    \label{37Ar}
\end{figure}

\begin{figure}[h]
    \centering
    \includegraphics[scale=0.8]{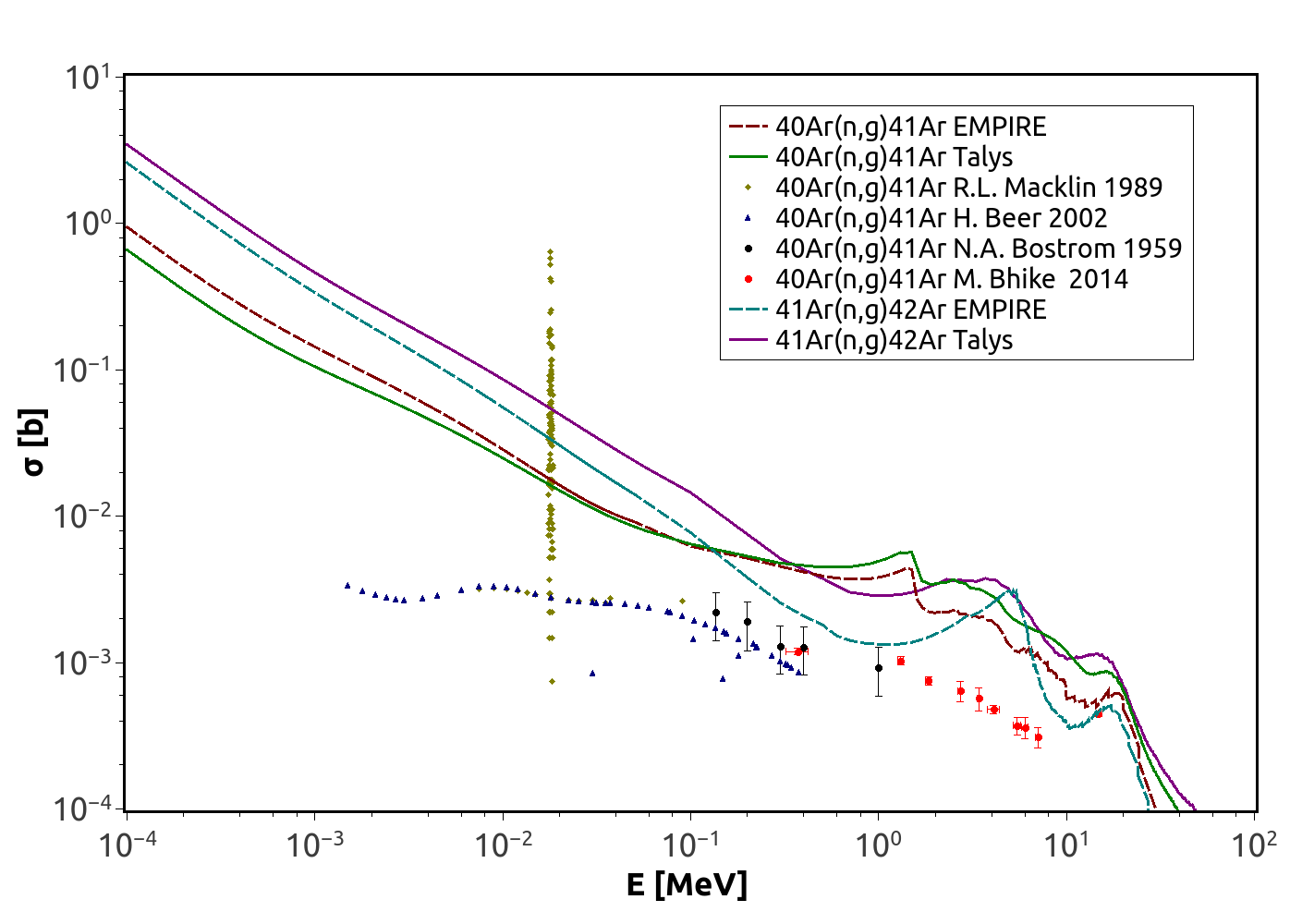}
    \caption{The energy dependence of the cross sections for the $^{40}$Ar(n,$\gamma$)$^{41}$Ar and $^{41}$Ar(n,$\gamma$)$^{42}$Ar reactions.}
    \label{42Ar}
\end{figure}

\begin{figure}[h]
    \centering
    \includegraphics[scale=0.8]{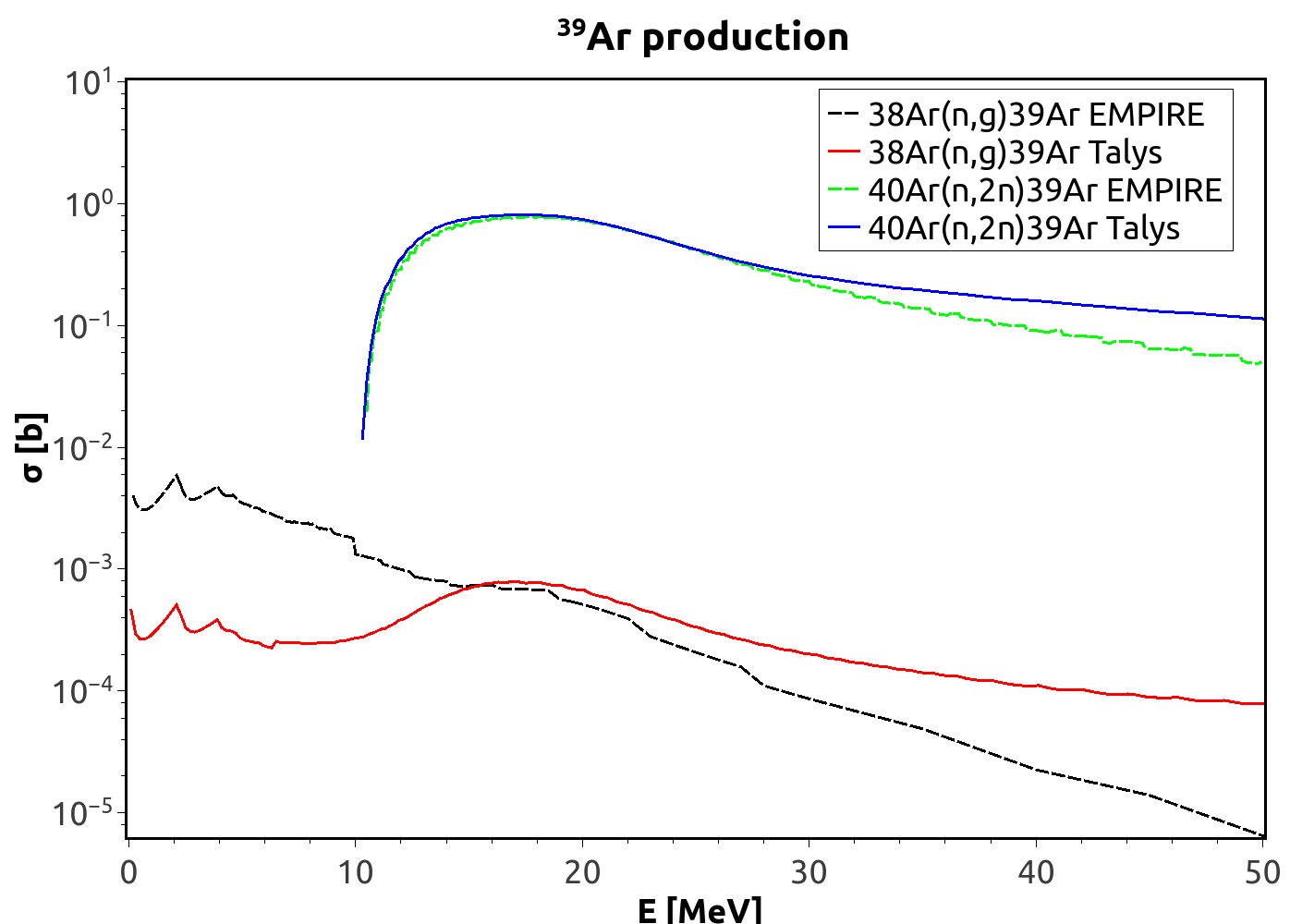}
    \caption{The energy dependence of the cross sections for the $^{40}$Ar(n,2n)$^{39}$Ar and $^{38}$Ar(n,$\gamma$)$^{39}$Ar reactions.}
    \label{39Ar}
\end{figure}

 The isotope $^{37}$Ar is produced in the reactions $^{40}$Ar(n,4n)$^{37}$Ar, $^{36}$Ar(n,$\gamma$)$^{37}$Ar or $^{38}$Ar(n,2n)$^{37}$Ar. The last long lived radioactive isotopes can be produced in the two steps reactions $^{40}$Ar(n,$\gamma$)$^{41}$Ar and thus $^{41}$Ar(n,$\gamma$)$^{42}$Ar. The problem of the production of radioactive isotopes as radioactive background of argon was considered previous by Parvu et. al. \cite{Parvu:2017xde} and R. Saldanha  et. al. \cite{Saldanha:2019zpu}. Unfortunately, for most of the reactions induced by neutrons is argon there is insufficient data and the few models available do not agree with each other. Estimates and experimental measurements (when they exist) of $^{39}$Ar, $^{37}$Ar and $^{42}$Ar production cross sections for the energy range of interest for the present paper are presented in the Figures \ref{37Ar}, \ref{42Ar}, \ref{39Ar}. 


\begin{figure}[h!]
    \centering
    \includegraphics[scale=0.25]{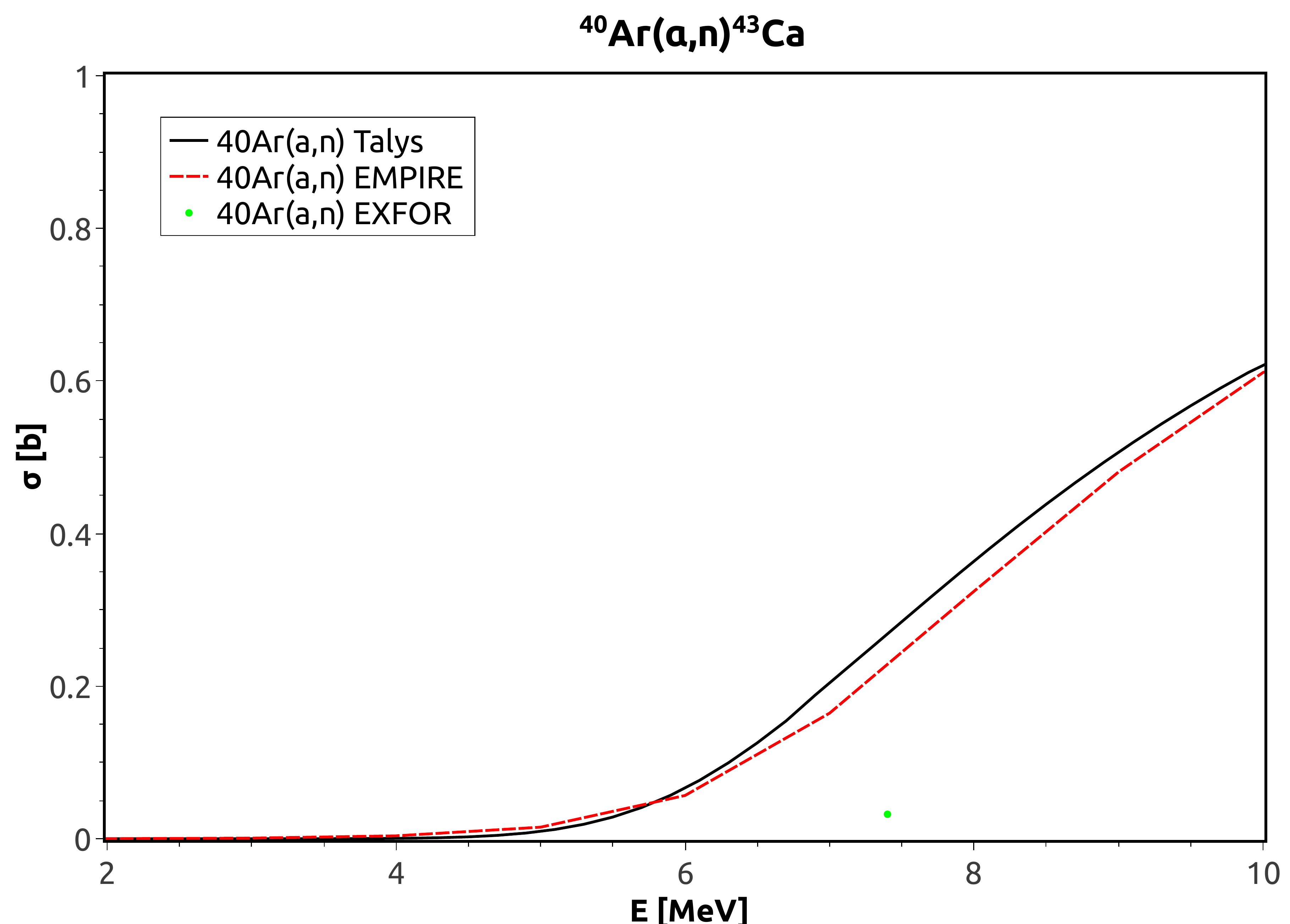}
    \caption{The energy dependence of the cross sections for the $^{40}$Ar($\alpha$,n)$^{43}$Ca reaction.}
    \label{Ar40nalpha}
\end{figure}

 The last calculation considers ($\alpha$,n) processes in $^{40}$Ar with the production of neutrons. For energies of $\alpha$ particles up to 10 MeV the two codes considered in this paper are in remarcable agreement. Unfortunately, only one experimental point is available in EXFOR database which is in disagreement with the calculations. The results are presented in Figure \ref{Ar40nalpha}. The energy dependence of the cross sections predicted by Talys and EMPIRE are in a good agreement. Unfortunately, the values are in major discrepancy with the single experimental point.

A supplementary background is expected due to muon spallation. A detailed study was published in Ref. \cite{Zhu:2018rwc}  for argon, considering the production of unstable isotopes, sources for later beta background. 

Recently, Saldanha and coworkers \cite{Saldanha:2019zpu} have done the first experimental measurements of $^{39}$Ar and $^{37}$Ar production rates from cosmic neutrons interactions and their results will be used to discriminate between the prediction of the cross sections obtained using different codes.

\section{Neutron production rates of radioactive isotopes of argon}

\subsection{Daily interaction rates of cosmic neutrons with argon}

\begin{figure}[h]
    \centering
    \includegraphics[scale=0.5]{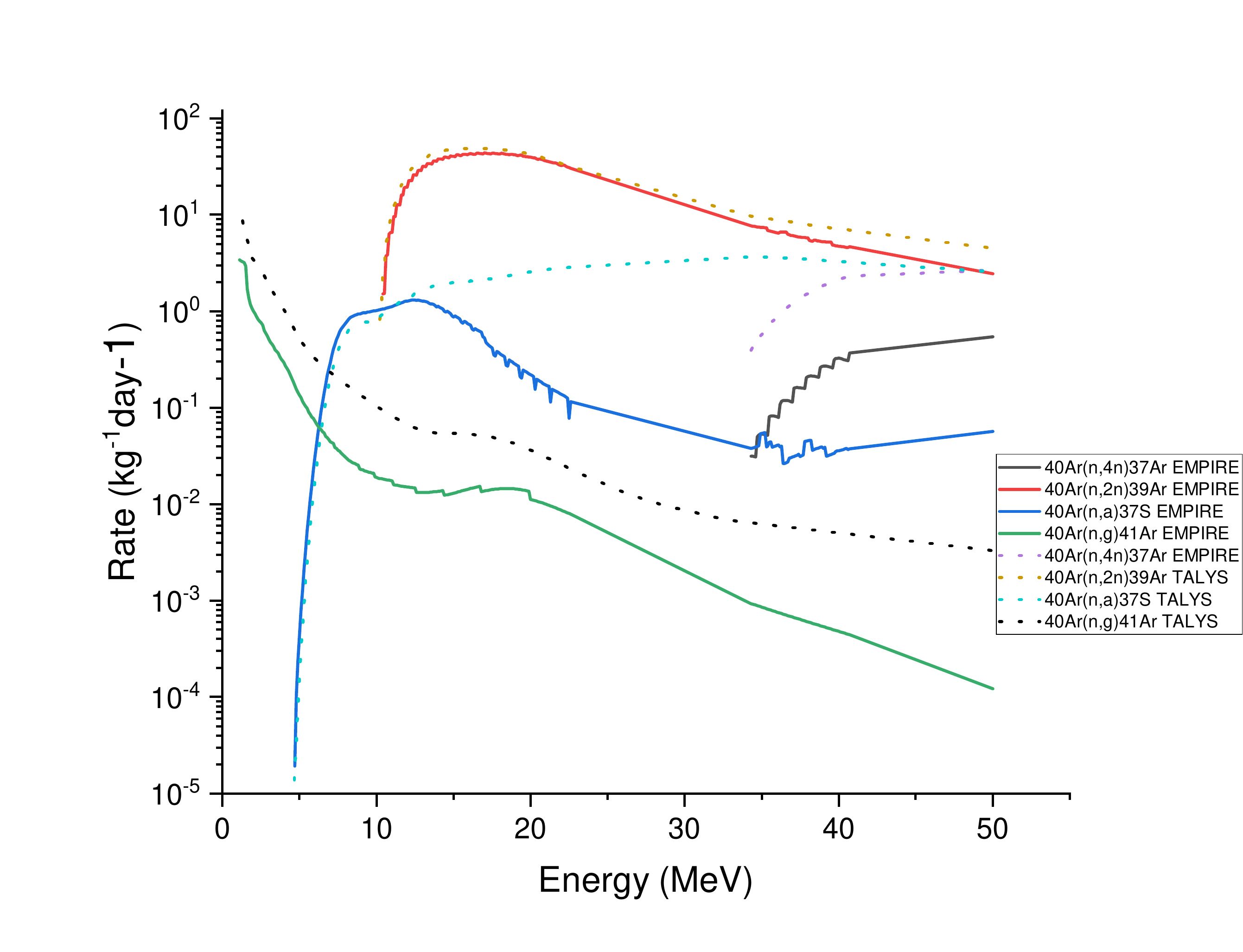}
    \caption{Daily rates.}
\label{40Ar_rates}
\end{figure}

In this analysis rates for the following reactions (n,2n), (n,4n), (n,p), (n,$\alpha$) and (n,$\gamma$) in $^{40}$Ar are calculated using the cross sections obtined with Talys and EMPIRE codes. These results are presented in Figure \ref{40Ar_rates}.

Maximum rates are obtained for $^{40}$Ar(n,$\gamma$)$^{41}$Ar and $^{40}$Ar(n,2n)$^{39}$Ar with values between 10 to 100 kg$^{-1}$ day$^{-1}$ for rapid neutrons, with energies of 10-15 MeV. For other reactions the rates are at least one order of magnitude smaller. Unfortunately, great discrepancies are obtained between the two sets of data. (n,4n) process has a high energy threshold and this behaviour can be observed both in Talys and EMPIRE calculations.

During the operation time of NP04 detector, the flux from cosmic neutrons must be supplemented with the contribution of the secondary neutrons produced by the beam line.

\subsection{Isotopes produced by major elements component of dectector's external structure}

The presence of the external structure made of stainless steel generates a supplementary radioactive background due to the radioactive isotopes produced in iron, chromium and nickel. (n,2n), (n,p) and (n,$\alpha$) processes are investigated. A short analysis of the produced isotopes are presented in Table \ref{reaction_rates_steel}. Data is extracted from \cite{atomkaeri}.

\begin{table}[h]
\begin{center}
\begin{tabular}{ |c|c|c|c|c|c| }  
 \hline
Isotope  & Decay mode & Decay energy [MeV] & Half-life\\
 \hline
 $^{55}$Fe & electron capture & 0.231 & 2.744 y\\ 
 \hline
$^{56}$Mn & $\beta^-$ & 3.695 & 2.5789 h\\ 
 \hline
$^{53}$Cr & stable & - & -\\
 \hline
 $^{51}$Cr & electron capture & 0.753 & 27.7 d\\ 
 \hline
$^{52}$V & $\beta^-$ & 3.976 & 3.743 min\\ 
 \hline
$^{49}$Ti & stable & - & -\\
 \hline
 $^{57}$Ni & electron capture & 3.264 & 35.6 h\\ 
 \hline
$^{58}$Co & electron capture & 2.307 & 70.86 d\\ 
 \hline
\end{tabular}
\caption{Isotopes produced by cosmic neutrons in major elements of stainless steel.}
\label{reaction_rates_steel}
\end{center}
\end{table}

\begin{figure}[h!]
 \begin{subfigure}{.5\textwidth}
 \centering 
  \includegraphics[scale=0.35]{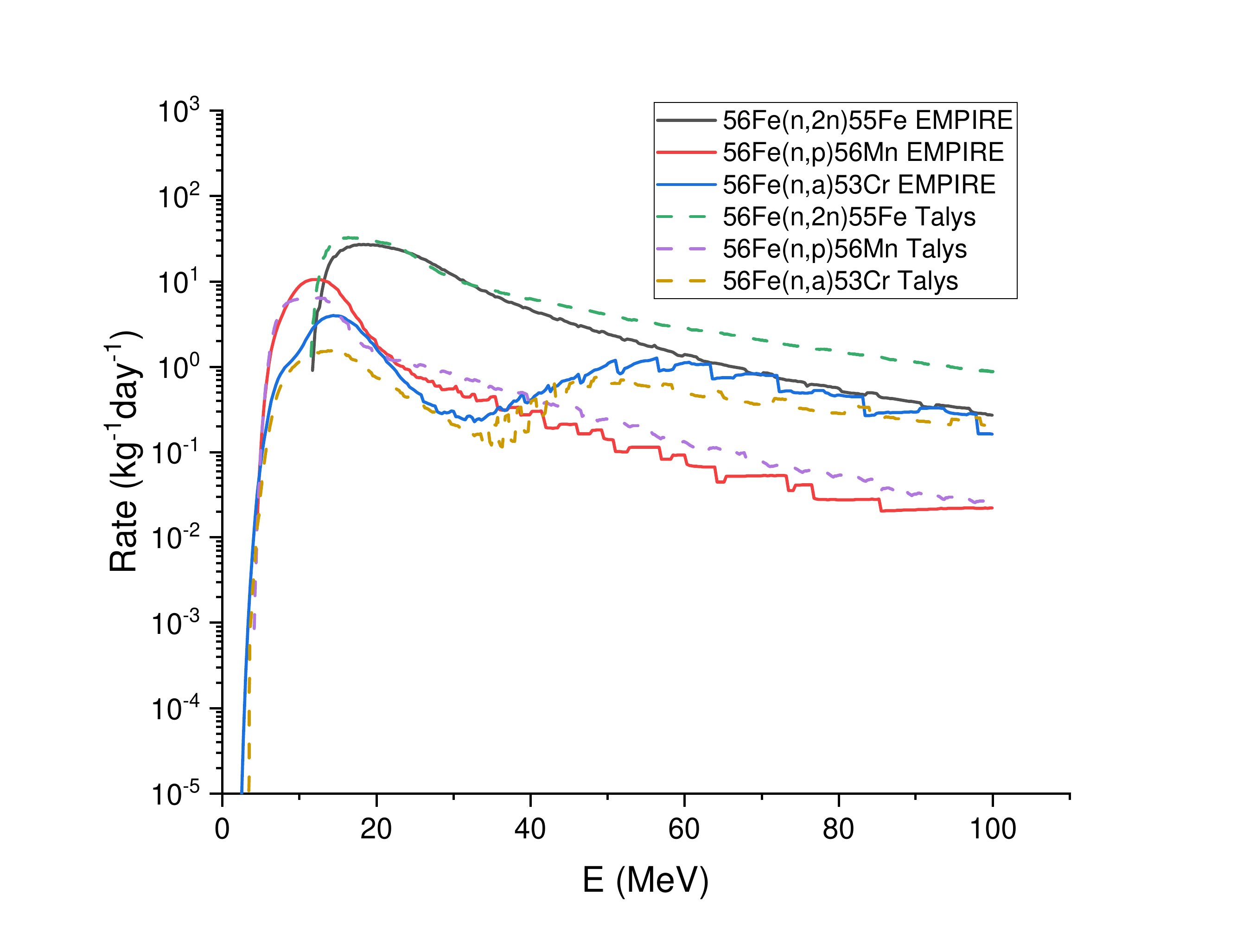}
    \caption{}
    \label{Fe_rate}
 \end{subfigure}
 \begin{subfigure}{.5\textwidth}
\centering  
  \includegraphics[scale=0.35]{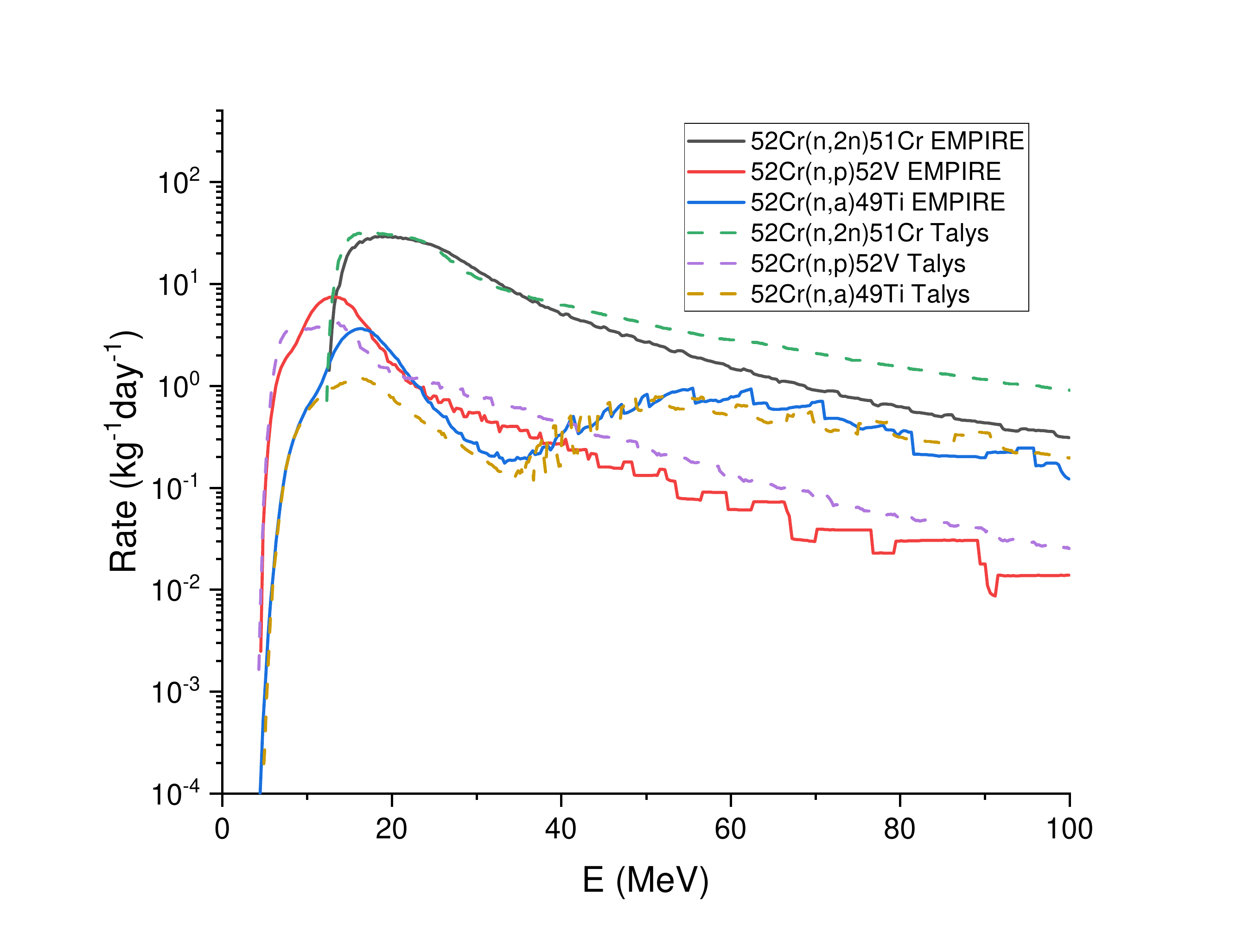}
    \caption{}
    \label{Cr_rate}
   \end{subfigure}

   \centering   
   \begin{subfigure}{.5\textwidth}
\centering    
    \includegraphics[scale=0.35]{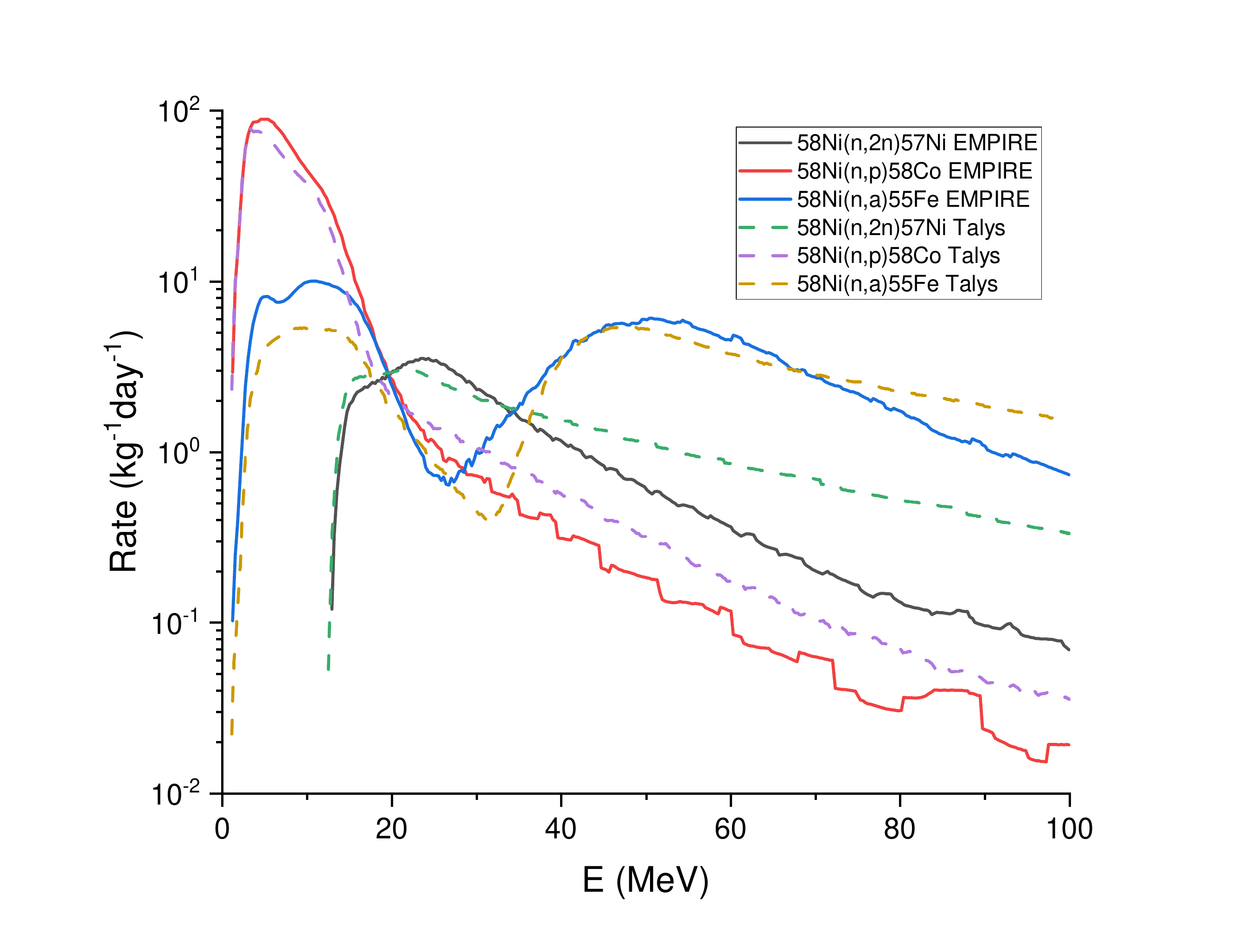}
    \caption{}
    \label{Ni_rate}
    \end{subfigure}
    \caption{a) Rates for (n,2n), (n,p) and (n,$\alpha$) processes in $^{56}$Fe. b) Rates for (n,2n), (n,p) and (n,$\alpha$) processes in $^{52}$Cr. c) Rates for (n,2n), (n,p) and (n,$\alpha$) processes in $^{58}$Ni.}
    \label{Rate}
\end{figure}

From the analysis of isotopes produced by cosmic neutrons two are stable ($^{53}$Cr,$^{49}$Ti), butin this case the presence of the $\alpha$ particles will induce reactions on $^{40}$Ar, in particular the ($\alpha$,n) process. All the other produced isotopes ($^{55}$Fe, $^{56}$Mn, $^{51}$Cr, $^{52}$V, $^{57}$Ni and $^{58}$Co) are electron emitters and except $^{55}$Fe can emit photons via Cherenkov effect and is a potential source of background for scintillating emission.

In all the calculated rates the results are in a relatively good agreement with both of the used codes. In Figures \ref{Fe_rate}, \ref{Cr_rate}, \ref{Ni_rate} are presented the obtained results. 

In a previous paper, Zhang et al \cite{Zhang:2016rlz} evaluated the cosmogenic production in some materials, including also stainless steel using Geant4 and ACTIVIA. Their results agree with our calculations. 

\section{Summary and conclusions}

 In this work we have investigated the radioactive background important for the ProtoDUNE-DP as next generation of liquid argon detectors. Cosmic muons and neutrons, as well as neutrons produced by beam interaction with the components of the NP02/ProtoDUNE-DP detector were taken into account. Using the available fluxes in literature for muons and neutrons from cosmic origin, as well as the predicted neutron flux produced by H4-VLE beam in the vicinity of the NP-02-DP detector, the daily doses inside of the cryostat were simulated considering simplified but realistic elements of the detector in the frame of FLUKA code. The simulated results put in evidence the dominant, non permanent contribution from beam-induced neutrons.\newline
Simple estimations using analytical approximations of the neutron yield per muon suggest that muons do not contribute significantly to increasing of the neutron flux for the energy range between 15 and 380 GeV and this class of processes are irrelevant in this case. Similar conclusions are obtained considering muon capture in argon and in the dominant elements of membrane of the cryostat that is in direct contact with the liquid argon.\newline
 Production mechanisms of unstable argon isotopes or other radioactive nuclei in bulk of LAr are obtained, presented and discussed as cross sections for considered processes. These results put in evidence the energy ranges where Talys and EMPIRE codes give similar results, regions with major discrepancies and where it exists, if are consistent with the experimental data.\newline
The calculations were made also for iron, nickel and chromium, elements which appear in the composition of stainless steel. Reactions induced by neutrons were considered: $^{56}{\rm Fe}(n,2n) ^{55}{\rm Fe}$, $^{56}{\rm Fe}(n,p) ^{56}{\rm Mn}$, $^{56}{\rm Fe}(n,\alpha) ^{53}{\rm Cr}$, $^{58}{\rm Ni}(n,2n) ^{57}{\rm Ni}$, $^{58}{\rm Ni}(n,p) ^{58}{\rm Co}$, $^{58}{\rm Ni}(n,\alpha) ^{55}{\rm Fe}$, $^{52}{\rm Cr}(n,2n) ^{51}{\rm Cr}$, $^{52}{\rm Cr}(n,p) ^{52}{\rm V}$ and $^{52}{\rm Cr}(n,\alpha) ^{49}{\rm Ti}$.\newline
 There are very few experimental measurements for the interactions of neutrons, proton and alpha particles with $^{40}$Ar and other of its isotopes. In the absence of these measurements, calculations done using Talys and EMPIRE codes represent a support for further studies.

\bibliography{references}

\acknowledgments
The authors warmly thank the Background Working Group of the DUNE collaboration for the very useful discussions during several meetings. Special thanks to Juergen Reichenbacher and Vitaly Kudryavtsev. Many thanks to John Beacom for his constructive comments and suggestions about the manuscript. Thanks to Mark Messier for his patience in our interaction with the Author's Publishing Committee.

This work was performed with the financial support of the Research Institute of the University of Bucharest (ICUB) and PNCDI III 2015-2020, Programme 5, Module CERN-RO, contract no. 2/2020.



\end{document}